\documentstyle[12pt]{article}

\def\bold#1{\setbox0=\hbox{$#1$}%
      \kern-.025em\copy0\kern-\wd0
      \kern.05em\copy0\kern-\wd0
      \kern-.025em\raise.0433em\box0 }

\newcommand{\beq}{\begin{equation}}
\newcommand{\eeq}{\end{equation}}
\newcommand{\beqn}{\begin{eqnarray}}
\newcommand{\eeqn}{\end{eqnarray}}
\newcommand{\redexpval}[3]
{\left\langle#1\left|\left|#2\right|\right|#3\right\rangle}
\newcommand{\toref}[1]{\mbox{(\ref{#1})}}

\begin{document}
\baselineskip 8mm
\begin{titlepage}
\begin{flushright}
Revised Version: April 27, 1997\\
Submitted to {\em Physical Review Letters}
\end{flushright}
\begin{center}
\vspace*{1.5cm}
{\large\bf Competition Between Particle-Hole and Particle-Particle 
           Correlations in Forbidden Electron Capture: the Case of $^{123}$Te}
\vskip 1cm
{M. Bianchetti$^{a}$, M.R.Quaglia$^{a}$, G. Col\`o$^{a}$, P.M.Pizzochero$^{a}$, 
R.A.Broglia$^{a,b}$, P.F.Bortignon$^{a}$}
\vskip .2cm
{\it $^a$ Dipartimento di Fisica, Universit\`a di Milano, \\
     and \\
     INFN, Sezione di Milano, \\
     Via Celoria, 16, I-20133 Milano, Italy,\\
     E-mail: Name@mi.infn.it\\
     $^b$ The Niels Bohr Institute, University of Copenhagen, \\
     2100 Copenhagen, Denmark}

\end{center}
\vskip 1cm
\begin{center}
{\bf ABSTRACT}\\
\begin{quotation}
The K-electron capture half-life of $^{123}$Te has been recently measured to be
$t^K_{\rm exp}=2.4\times 10^{19}$ yr, and constitutes the longest half-life
ever measured in a single $\beta$-transition of any nuclear species. We have
calculated this second unique forbidden transition within the framework of the
proton-neutron quasi-particle random phase approximation, making use of
Skyrme-type effective interactions. A strong cancellation effect between
particle-hole and particle-particle correlations is found. The model, without
any renormalization of the force, provides a lower limit for the K-electron
capture half-life of $\sim 10^{17}$ yr, which unambiguously rules out the old
experimental values of $10^{13} - 10^{14}$ yr. A few percent increase of the 
particle-particle matrix elements of the Skyrme interaction allows to reproduce 
the experimental findings. 

\end{quotation} 
\vspace{0.5cm}
\end{center}
\end{titlepage}
%
%
Experimental evidence for a strongly suppressed second unique forbidden
electron capture (EC) process in $^{123}$Te was recently obtained at the Gran
Sasso Underground Laboratory, making use of sophisticated calorimetric
techniques \cite{Fiorini}. The resulting half-life for K-electron capture was
reported to be $t^K_{\rm exp}=2.4 \times10^{19}$ yr, a value which is in
contradiction with both the value of $10^{13} - 10^{14}$ yr reported in the
tables and the systematics of these processes \cite{nds}. To our knowledge,
this is the longest half-life ever measured in a natural single
$\beta$-transition of any nuclear species. Since atomic and $Q$-value effects
account only for a factor of $\sim 10^{16}$ in the $K$ half life, the remaining
factor of $\sim 10^3$ testifies to the action of an extremely strong
cancellation mechanism between particle-hole and particle-particle correlations
\cite{footnote1}. These facts make a theoretical calculation worth doing. 
\par
The interplay between particle-hole and particle-particle 
contributions to nuclear matrix elements is found in a wide range of nuclear 
phenomena (cf. e.g. Refs. \cite{Mot,Bes,Bro} and refs. therein). 
Systematic studies of these effects in channels different 
from the p-n channel, which are as a rule easier to deal with, 
have indicated the importance of using a realistic configuration space 
to calculate the wavefunctions describing the initial and final 
states connected by the transition. 
In keeping with these facts, in what follows we shall calculate the 
half-life of the second unique forbidden decay of $^{123}$Te 
using as input various experimental data obtained from one-nucleon transfer 
experiments. The microscopic structure of the initial and final configurations
have been worked out in the proton-neutron quasi-particle random phase
approximation (pn-QRPA), making use of Skyrme-type interactions. 
Since the Skyrme parameters \cite{VauBei} are fixed mainly to reproduce 
bulk and particle-hole-like nuclear properties in calculations carried 
out in a limited single-particle subspace, one can hardly trust
their particle-particle matrix elements, nor can one use the purely
geometric Pandya transformation directly. For this reason, a 
multiplicative factor G$_{pp}$ is introduced, which renormalizes the
particle-particle matrix elements calculated via the Pandya transformation
(cf., e.g., Ref. \cite{vogel}). This parameter reflects a basic limitation
in the present understanding of the nuclear interaction.
\par
The results of our calculation provide a lower limit for the K-electron capture
half-life, of the  order of $\sim 10^{17}$ yr. The value  $t^K_{\frac{1}{2}}=
1.3 \times 10^{17}$ yr is obtained if no renormalization of the Skyrme
particle-particle matrix elements is introduced (G$_{pp}=1$), while the
experimental half-life of Ref.~\cite{Fiorini} is reproduced with G$_{pp}=1.06$.
\par
The process under consideration is
\beq
^{123}\mbox{Te}\left(\mbox{\small{GS}}, J_i^{\pi_i}=\frac{1}{2}^+\right) +e^- 
\longrightarrow 
^{123}\mbox{Sb}\left(\mbox{\small{GS}}, J_f^{\pi_f}=\frac{7}{2}^+\right) +\nu_e,
\nonumber
\label{difficult}
\eeq
with $\Delta J=3$ and $\pi_i \pi_f=+1$. It falls into the second unique
forbidden class (see e.g. refs. \cite{grotz,EC,BM1}). Following the theory of
EC \cite{EC,BuhringSchulke}, this type of forbidden transitions can only
proceed through capture from K, L, M atomic shells. 
Neglecting corrections induced by the 
strong interaction, the partial decay constants are given by 
(with $\hbar=c=m_e=1$)
\beq
\lambda_{x}=\frac{\ln 2}{t^x_{\frac{1}{2}}}=
\frac{8}{15\pi}G^2_\beta 
\frac{p_x^{2(k_x-1)}q_x^{8-2k_x}}{(2k_x-1)!(7-2k_x)!}\,\beta^2_x B_x
\times B(\hat{T}^+_{3,2,1};\,i\rightarrow f),
\label{partialdecayconstant}
\eeq
with $k_x=1$ for K, L$_{1,2}$, M$_{1,2}$ atomic orbitals, $k_x=2$ for L$_{3}$,
M$_{3,4}$, and $k_x=3$ for M$_{5}$. The quantity $G_\beta$ is the weak
interaction coupling constant involved in $\beta^-$ decay processes, $p_x$ is
the momentum of the electron captured from shell $x$ and $q_x$ is the momentum
of the emitted neutrino. Neglecting nuclear recoil, we have $q_x=Q_V-E_x$,
where $E_x$ is the binding energy of atomic shell $x$, while $Q_V=0.0513$ MeV
($=0.1$ in the units introduced above) \cite{nds} is the very small Q-value of
the reaction.  It is worth stressing  that  the smallness of this  Q-value
leads to the emission of a very low energy neutrino ($q_x \ll 1$), which
accounts for both the strong suppression of the decay constants and for the
prominent role played by capture processes from higher atomic shells. The
quantity $\beta_x$ is the Coulomb amplitude for the shell $x$, that is the
probability amplitude to find a $x$-electron inside the nucleus $^{123}$Te. The
quantity $B_x$ is a correction factor taking into account the atomic exchange
and overlap effects associated with the $x-$electron \cite{footnote2}. 
\par
Nuclear structure enters the calculation of $\lambda_x$ through the quantity 
\beq
  B(\hat{T}^+_{3,2,1};\;i\rightarrow f)=\left(\frac{g_A}{g_V}\right)^2
  \frac{1}{2J_i+1}\left|\redexpval{f}{\hat{T}^+_{3,2,1}}{i}\right|^2 ,
\label{transprob}
\eeq
that is, the reduced transition probability induced by the charge-exchange
spin-flip quadrupole operator
$\hat{T}^+_{3,2,1}=r^2\vec{Y}_{3,2}\cdot\vec{\sigma}\,\tau_+$ 
(we follow the notation of Ref. \cite{EC}) acting between the 
initial and final nuclear states.
In keeping with the well known quenching effect 
associated with Gamow-Teller strength found in p-n reactions \cite{Ost}, 
we shall use the effective value $g_A/g_V = 1$
for the ratio between axial and vector weak coupling constants \cite{BW}. 
Making use of the experimental results of Ref. \cite{Fiorini} and of Eq. 
\toref{partialdecayconstant}, one can extract an {\em empirical} nuclear 
matrix element, which, reverting to standard units, is 
\beq
\left|\redexpval{f}{\hat{T}^+_{3,2,1}}{i}\right|^2
=1.8\times 10^{-3}\mbox{ fm}^4.
\label{empirical}
\eeq
Using again Eq.~\toref{partialdecayconstant} to scale the atomic 
contribution, one can deduce empirical 
estimates for the half-lives of EC from higher atomic shells. 
These are reported in column 2 of Table 1. 
We stress that these results depend only on rather general assumptions 
concerning standard EC theory, and are independent of any nuclear
structure model once the matrix element in Eq. \toref{empirical} is known.
\par
One-nucleon transfer experiments \cite{078} indicate that the main component
(78\%) of the wave function describing the ground state of 
$^{123}_{52}$Te$_{71}$ is a pure $^{122}_{52}$Te$_{70}$ 
even-even core plus an odd $3s_{\frac{1}{2}}$ neutron. 
Within this model space, the only configuration in the daughter 
nucleus $^{123}_{51}$Sb$_{72}$ which gives a sizable contribution 
to the transition matrix element appearing in Eq. \toref{transprob} 
corresponds to the coupling to $J_f^{\pi_f}=7/2^+$ between 
the lowest $J^{\pi}=3^+_1$ excited state of $^{122}_{51}$Sb$_{71}$ core 
and the $3s_{\frac{1}{2}}$ neutron. 
This configuration has been measured to be present in the ground 
state wavefunction of the daughter 
nucleus with a probability of 16\% \cite{016}. 
Other configurations which can contribute to the 
nuclear matrix element in Eq. \toref{transprob} involve 
either higher order forbidden transitions or contain hindrance factors. 
\par
Within this scenario, the decay process
in Eq. \toref{difficult} is equivalent to the transition 
\beq
^{122}\mbox{Te}\left(\mbox{\small{GS}},J_i^{\pi_i}=0^+\right) +e^- 
\longrightarrow
^{122}\mbox{Sb}\left(J_f^{\pi_f}=3^+_1\right) +\nu_{e}.
\label{simple}
\eeq
The contribution of the spectator neutron configuration 
$\nu 3s_{\frac{1}{2}}$ may be factored out  
in terms of the matrix element of Eq. \toref{transprob} as follows 
(cf. e.g Ref.\cite{BM1})
\beqn
\left|\redexpval
   {\left[\left(^{122}\mbox{Sb}, J^\pi = 3_1^+\right)\otimes 
   \left(\nu 3s_{\frac{1}{2}}\right)\right]_{\frac{7}{2}^+}}
  {\hat{T}^+_{3,2,1}}
  {\left[\left(^{122}\mbox{Te}, J^\pi=0^+\right)\otimes 
   \left(\nu 3s_{\frac{1}{2}}\right)\right]_{\frac{1}{2}^+}}\right|^2
  \nonumber\\ 
=\frac{8}{7}\left|\redexpval{^{122}\mbox{Sb},J^{\pi}=3_1^+}
     {\hat{T}^+_{3,2,1}}{^{122}\mbox{Te},J^{\pi}=0^+}\right|^2.
\label{BEden}
\eeqn
The problem has thus been reduced to the calculation of the
matrix element of the operator $\hat{T}^+_{3,2,1}$, 
from the even-even system $^{122}$Te to the lowest $3^+_1$
state in $^{122}$Sb. To make sure that violations of the Pauli principle do 
not make our model unreliable, we have checked numerically that the inclusion 
or exclusion from our basis of the quasi-particle configurations involving the 
spectator state $\nu 3s_{\frac{1}{2}}$ does not change the results for the 
transition probabilities significantly.
\par
To do this, the mean-field of $^{122}$Te is described in terms 
of a Woods-Saxon potential parameterized according to Ref. \cite{BM1}. 
The spin-orbit strengths are adjusted to reproduce the correct
single particle level order \cite{ordering}.
Pairing correlations are taken into account in the BCS approximation, 
making use of experimental proton and neutron pairing gaps 
deduced from the experimental odd-even mass difference \cite{BM1}.
Some quasi-particle energies and occupation factors are 
substituted with their corresponding experimental values, where available 
\cite{exp}.
As nuclear residual interaction, we use the set of Skyrme parameters, SGII,
which was specially designed to reproduce spin-isospin properties of nuclei
\cite{sg1}.  
Both particle-hole and particle-particle matrix elements, related via the
Pandya transformation, are introduced,  and the latter are renormalized via the
multiplicative parameter $G_{pp}$. These interactions are diagonalized in the
standard pn-QRPA scheme within a proton-neutron two quasi-particle basis. The
$3^+$ spectrum of the $\beta^+$ daughter nucleus $^{122}$Sb is thus obtained,
and the squared transition matrix element defined in Eq. \toref{BEden} is
calculated as a function of the parameter G$_{pp}$. The introduction of the
experimental data obtained from stripping and pick-up reactions
\cite{078,npickup,pstripping,ppickup} and, in particular, of the quasi-proton
energies reduces the value of the transition matrix element of about one order
of magnitude and drastically changes the features of its dependence on the
strength G$_{pp}$ of the particle-particle interaction. Finally, we average the
$3^+$ strength associated to the few states found within $0.2$ MeV of
excitation energy, and we multiply this strength by the factor $0.78 \times
0.16$ discussed above. 
\par
The results for the half-lives are collected in Table 1.   From the
calculation, it turns out that sufficiently large values of G$_{pp}$ lead to a
divergence of the matrix element under study, a result typical of a phase
transition within the pn-QRPA formalism and well known as Thouless or RPA
instability \cite{rpnqrpa}. We shall thus disregard results associated with
this region as non-physical. Within the physical region the transition strength
varies in a range from $0$ to a maximum of $\sim 3$ fm$^4$. The zero value
corresponds to an {\em exact} cancellation between particle-hole and
particle-particle correlations (cf. e.g. Ref. \cite{vogel}). As a consequence,
we may conservatively extract only lower limits for EC half-lives. These are
shown in column 3 of Table \ref{tab1}. In particular, we point out the result
for the lower limit of the K-capture half-life, $\sim 10^{17}$ yr, which lies
right between the old and the new experimental data. 
\par 
In column 4 of Table 1 we report the half-lives corresponding 
to G$_{pp}=1$, that is, with no renormalization of 
the Skyrme particle-particle matrix elements. These values are close to the
lower limit discussed before. It can be seen from the table that the K
half-life in this case is $1.3 \times 10^{17}$ yr, so that two orders of
magnitude still separate the theoretical and the experimental half-lives. These
can be matched introducing a slight renormalization of the particle-particle
correlations, namely using the value G$_{pp}=1.06$. 
\par 
In conclusion,  a strong cancellation effect between particle-hole and
particle-particle correlations  appears already at the level of the
pn-QRPA with Skyrme-type effective interactions.  Theory can account
for the experimental findings  introducing a $6\%$ renormalization of the
particle-particle matrix elements. Estimates with G$_{pp}=1$ provide anyway a
lower limit, which can clearly rule out the results of previous observations. 
\\
\par
Discussions with F. Alasia, N. Giovanardi, B. Lauritzen, 
B. R. Mottelson, E. Ormand, N. Blasi and with the authors of Ref. 
\cite{Fiorini} are gratefully acknowledged. 
We also thank the referees for suggestions concerning 
the experimental data about the structure of the wavefunctions 
which describe the states connected by the transition.

\vfill
\newpage

\begin{table}[hbt]
\begin{center}
\begin{tabular}{|c|c|c|c|}
\hline\hline
$x$   & 
\multicolumn{3}{c|}{$t_{\frac{1}{2}}^x\;(yr)$} \\
\cline{2-4}
& empirical & lower limit  & G$_{pp}=1$ \\ 
\hline
K     & 2.4$\times 10^{19}$ & $ \sim 10^{17}$ 
& $1.3\times 10^{17}$ \\ \hline
L     & 6.1$\times 10^{16}$ & $ \sim 10^{14}$ 
& $3.2\times 10^{14}$ \\ \hline
M     & 1.6$\times 10^{17}$ & $ \sim  10^{15}$ 
& $8.6\times 10^{14}$ \\ \hline
Total & 4.3$\times 10^{16}$ & $ \sim  10^{14}$ 
& $2.3\times 10^{14}$ \\ \hline\hline
\end{tabular}
\end{center}
\caption{Decay half-lives for electron capture from different atomic orbitals,
expressed in years. Column 2: empirical half-lives deduced from K-capture
experimental data. The first entry comes from the experiment reported in Ref.
\protect\cite{Fiorini}, while the other entries are derived as described in the
text. These results can be reproduced theoretically by setting G$_{pp}=1.06$.
Column 3: theoretical lower limits. Column 4: theoretical results obtained with
G$_{pp}=1$.} 
\label{tab1}
\end{table}


\newpage 

\begin{thebibliography}{99}
\bibitem{Fiorini} A. Alessandrello  {\em et al.}, Phys. Rev. Lett. 
                  {\bf 77}, 3319 (1996).
\bibitem{nds} S. Ohya, T.Tamura, Nuclear Data Sheets {\bf 70}, 531 (1993).
\bibitem{footnote1} To be noted that the A=128,130 isotopes of 
  Tellurium are $2\nu\beta\beta$ emitters and good candidates for 
  $0\nu\beta\beta$. From a nuclear structure point of view, the study of 
  both EC and $\beta\beta$ processes can shed light on the mechanism which 
  is at the basis of the cancellation between particle-hole and  
  particle-particle correlations (cf.~\cite{vogel,grotz} and refs. therein). 
\bibitem{vogel} P. Vogel {\em et al.}, Phys. Rev. Lett. {\bf 57}, 3148 (1986).
\bibitem{grotz} G. Grotz and H.V. Klapdor-K., {\em The Weak Interaction in
                Nuclear, Particle and Astrophysics\/}, Adam Hilgher, 
                Bristol (1990).
\bibitem{Mot} B. R. Mottelson, {\em Int. School of Physics E. Fermi on 
	Nuclear Structure}, Ed. G. Racah, Academic Press, New York, 44 (1960).
\bibitem{Bes} D. R. Bes, R. Sorensen, {\em Adv. in Nucl. Phys.}, Vol. 2, 129,
              Ed. M. Baranger and E. Vogt, Plenum Press, New York (1969).
\bibitem{Bro} R. A. Broglia, O. Hansen, C. Riedel, {\em Adv. in Nucl. Phys.}, 
        Vol. 6, 287, Ed. M. Baranger and E. Vogt, Plenum Press, New York (1972).
\bibitem{VauBei} D. Vautherin, D.M. Brink, Phys. Rev. C {\bf 5}, 626 (1972); 
                 M. Beiner, H. Flocard, N. Van Giai, P. Quentin, 
                 Nucl. Phys. {\bf A238}, 29 (1975).
\bibitem{EC} W. Bambynek {\em et al.}, Rev. Mod. Phys. {\bf 49}, 77 (1977).
\bibitem{BM1} A. Bohr, B.R. Mottelson, {\em Nuclear Structure\/}, 
              Vol. I, W. A. Benjamin Reading Inc., Massachussets (1969). 
\bibitem{BuhringSchulke} W. B\"{u}rhing, L. Sch\"{u}lke, Nucl. Phys. 
                         {\bf 65}, 369 (1965).
\bibitem{footnote2} In Ref. \protect\cite{EC} the numerical values for 
        B$_2\ldots$B$_5$ are not reported. We have set them equal to 1, 
        as suggested by the systematics of the quoted values.
\bibitem{Ost} F. Osterfeld,     Rev. Mod. Phys. {\bf 64} (1992).
\bibitem{BW} B.A. Brown, B.H. Wildenthal, Atomic Data and Nuclear Data
             Tables, {\bf 33}, 347 (1985).
\bibitem{078} J. R. Lien {\em et al.},  Nucl. Phys. {\bf A253}, 165 (1975).
\bibitem{016} Hjort, Arkiv Fysik {\bf 33}, 183 (1967).
\bibitem{ordering} The order of single particle levels may be deduced
  from one-particle transfer reactions \cite{078,npickup,pstripping,ppickup}.
  In particular, we assumed a $(\pi 1g_{\frac{7}{2}})^2$ configuration
  for the two protons out of the $Z=50$ core in the 
  $^{123}_{52}$Te ground state \cite{pstripping}.
\bibitem{exp} Neutron quasi-particle energies in $^{123}$Te 
 are taken from Ref. \cite{078}. Quasi-proton energies in 
 $^{123}$I are given in Ref. \cite{pstripping}, 
 and have been referred to the ground state of 
 $^{123}$Te according to the scheme followed in Ref. \cite{078}.
 The experimental quasi-particle occupation factors in 
 $^{122}$Te are given in Refs. 
 \cite{078,npickup,pstripping,ppickup} and were renormalized in each channel 
  to sum up to the correct number of valence nucleons. 
\bibitem{sg1} N. Van Giai, H. Sagawa, Phys. Lett. B {\bf 113}, 119 (1982).
\bibitem{npickup} M. A. G. Fernandes, M. N. Rao, J. Phys. {\bf G10}, 1397 
            (1977); S. Gales {\em et al.}, Nucl. Phys. {\bf A381}, 173 (1982).
\bibitem{pstripping} J. R. Lien {\em et al.}, Nucl. Phys. {\bf A281}, 443 (1977).
\bibitem{ppickup} M. Conjeaud {\em et al.}, Nucl. Phys. {\bf A215}, 382
(1973). 
\bibitem{rpnqrpa} It is well known that this phenomenon is associated 
  with the appearance of a zero energy state, and is a signal
  of the failure of the standard pn-QRPA model, due to ground state
  correlations induced by the increasing particle-particle interaction.
\end{thebibliography}
\end{document}